# A novel approach for the automated segmentation and volume quantification of cardiac fats on computed tomography

É.O. Rodrigues [a,*], F.F.C. Morais [b], N.A.O.S. Morais [b], L.S. Conci [c], L.V. Neto [b], A. Conci [a]

[a] Department of Computer Science, Universidade Federal Fluminense (UFF), Rua Passo da Pátria 156, Niterói, Rio de Janeiro, Brazil
[b] Department of Internal Medicine and Endocrine Unit, Medical School and Hospital Universitário Clementino Fraga Filho, Universidade Federal do Rio de Janeiro (UFRJ), Rua Rodolpho Paulo Rocco, 255 – Cidade Universitária, Rio de Janeiro, Brazil
[c] Department of Specialized Medicine, Universidade Federal do Espírito Santo (UFES), Av. Marechal Campos, 1468 – Maruípe, Vitória, Brazil

## ARTICLE INFO

## ABSTRACT

The deposits of fat on the surroundings of the heart are correlated to several health risk factors such as atherosclerosis, carotid stiffness, coronary artery calcification, atrial fibrillation and many others. These deposits vary unrelated to obesity, which reinforces its direct segmentation for further quantification. However, manual segmentation of these fats has not been widely deployed in clinical practice due to the required human workload and consequential high cost of physicians and technicians. In this work, we propose a unified method for an autonomous segmentation and quantification of two types of cardiac fats. The segmented fats are termed epicardial and mediastinal, and stand apart from each other by the pericardium. Much effort was devoted to achieve minimal user intervention. The proposed methodology mainly comprises registration and classification algorithms to perform the desired segmentation. We compare the performance of several classification algorithms on this task, including neural networks, probabilistic models and decision tree algorithms. Experimental results of the proposed methodology have shown that the mean accuracy regarding both epicardial and mediastinal fats is 98.5% (99.5% if the features are normalized), with a mean true positive rate of 98.0%. In average, the Dice similarity index was equal to 97.6%.

## 1. Introduction

The cardiac epicardial and mediastinal (also termed pericardial) fats are correlated to several cardiovascular risk factors

[1]. At the present, three techniques appear suitable for the quantification of these adipose tissues, namely: magnetic resonance imaging (MRI), echocardiography and computed tomography (CT). Although these modalities have been widely used in several studies in the literature [2–4], computed

* Corresponding author. Tel.: +55 21986161930.
E-mail address: erickr@id.uff.br (É.O. Rodrigues).

tomography provides a more accurate evaluation of fat tissues due to its higher spatial resolution when compared to ultrasound and MRI [5]. In addition, CT is also widely used for evaluating coronary calcium score [4].

The automated quantitative analysis of epicardial and mediastinal fats may add a prognostic value to cardiac CT trials, ensuring an improvement on its cost-effectiveness. Besides, that automation diminishes the variability introduced by different observers. In fact, quantifying these data by direct user intervention is highly prone to inter and intra-observer variability. Thus, quantified samples may not be associated to a unified common sense of segmentation. Moreover, Iacobellis et al. [6] addressed the fact that the thickness of the epicardial fat and coronary artery disease, for instance, correlate independently of obesity. This evidence supports the individual segmentation and quantification of the adipose tissues rather than merely and simply estimating their volumes based on the overall fat of the patient.

### 1.1. *Contributions of this work*

In this work, we propose a novel methodology for automatically segmenting and consequently discriminating both the epicardial and mediastinal fats on cardiac CT images. To the extent of our knowledge, there is currently no unified method in the literature capable of both types of autonomous segmentations. On the entire extent of the proposed methods for cardiac fat segmentation [5,7–10], the addressed issue is specifically the segmentation of the epicardial fat, whereas no work has ever attempted to segment the mediastinal. Moreover, we resolve the segmentation issue diverging greatly with respect to the basis of the approach in relation to other works by employing an intersubject registration along with classification algorithms to produce the segmentation.

Summarily, this work contributes mainly to the field of visual computing by, namely: (1) proposing an accurate intersubject registration for cardiac CT images, (2) developing and analysing a robust hybrid similarity measure, applied within the registration procedure, (3) designing a new feature based on the Gaussian Kernel, (4) corroborating on the appliance of classification algorithms for image segmentation, (5) analysing the performance and accuracy of various classifiers for the problem, (6) creating a ground truth of cardiac fats available online and, mainly, by developing and evaluating (7) a unified and fully automatic segmentation methodology for both epicardial and mediastinal fats on cardiac images.

## 2. Literature review

The human heart is enclosed in the pericardium, a fibroserous sac comprising three concentric layers. The outermost layer is a densely fibrous, tough and inelastic structure, the fibrous pericardium. Inside the fibrous pericardium is the serous pericardium, which consists of two layers; the outer of these (which is firmly applied to the inner surface of the fibrous pericardium) is termed parietal layer. This layer is reflected around the roots of the great vessels to become continuous with the visceral layer (epicardium), which covers the internal surface

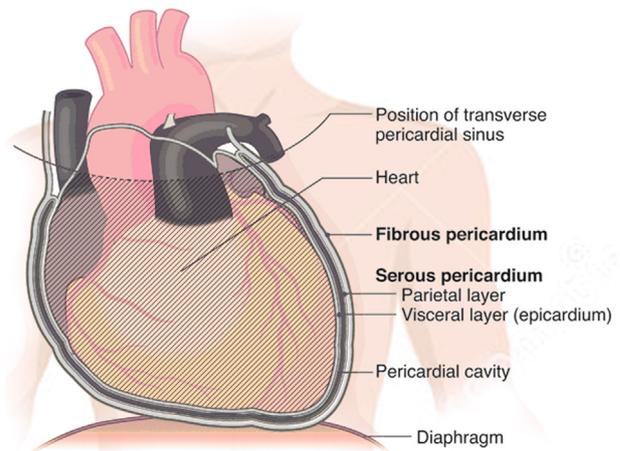

**Fig. 1 – Outermost layers of the heart.**

of the heart and is firmly applied to it [11]. Each one of these layers is depicted in Fig. 1.

Sacks et al. [12] define the epicardium or visceral layer of the pericardium as a population of mesothelial cells that migrate onto the surface of the heart from the area of the septum transversum (the embryological source of the diaphragm). Furthermore, they define that, in the normal adult, the epicardial fat is concentrated in the atrioventricular and interventricular grooves and along the major branches of the coronary arteries as well as, to a lesser extent, around the atria, over the free wall of the right ventricle and over the apex of the left ventricle. The authors also define pericardial fat as all the epicardial plus the paracardial fat and, consequently, define that paracardial is the fat located on the external surface of the parietal pericardium (also within the mediastinum). They also highlight that paracardial fats have been alternatively termed mediastinal fats in the literature. The mediastinal area is shown in Fig. 2.

Sicari et al. [2] define that cardiac fats can be distinguished between two types of deposits: (1) the epicardial, which they describe exactly with the same words as defined by Sacks et al. [12] and, (2) the pericardial, which they define as being the fat situated on the external surface of the parietal pericardium within the mediastinum (alternatively termed mediastinal or intrathoracic fat). Nevertheless, Rosito et al. [1], Dey et al. [10] and Nichols et al. [13], for instance, describe the pericardial fat as being any adipose tissue within the pericardial sac.

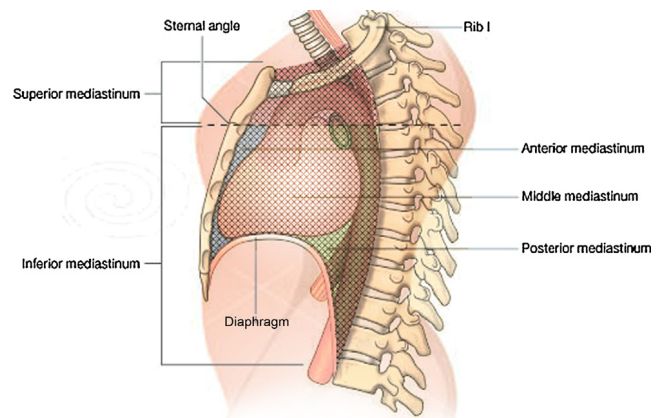

**Fig. 2 – Mediastinal space.**

However, Marwam et al. [14] addressed the fact that the terminology used to define fat deposits surrounding the heart in the current literature is diverse and, to some extent, confusing. The authors [14] state that the adipose tissue enclosed in the pericardial sac is frequently referred as pericardial fat. Despite of being a widely used definition, they state that the more accurate term would be epicardial fat, given its location on the internal epicardial surface of the heart [14]. Shahzad et al. [7] also mentioned this confusion regarding the terminology.

In summary, the majority of the published works [2,5,7–9,12,14–22] agree on the epicardial fat terminology as being correct for the fat contained within the epicardium and, therefore, also within the pericardium. On the other hand, the only work that has been out of that agreement on the epicardial fat definition is Mahabadi's et al. [23]. Nevertheless, there is an accentuated disagreement on the pericardial fat terminology amongst various works. In fact, some works [1,7,10,13,24,25] support the idea that the pericardial fat is merely the fat that is enclosed by the pericardium, which is analogous to the epicardial fat definition. Others [2,16,19,26–28] support that the pericardial fat terminology defines the adipose tissue located on the external surface of the parietal pericardium, within the mediastinum. Moreover, some works [12,20,29–31] even define the pericardial fat as being equivalent to all the adipose tissue within the mediastinum, including the pericardial or epicardial fats. Thus, due to the fact that the terminology for cardiac fats is currently not properly established in the literature, we define the fat located within the epicardium as epicardial, agreeing with the majority of published works. By following the same "first outer anatomical container" logic, we conclude that mediastinal is the best definition for the fat located on the external surface of the heart or of the fibrous pericardium. In other words, by taking Fig. 2 as reference, the mediastinal fat is located in the hachured area (within the anterior, middle and posterior mediastinum), and is mediastinal as long as it is not epicardial, i.e., as long as it is not located within the epicardium.

### 2.1. Adipose tissue and its associated health risks

Some studies associate the amount of epicardial fat to the progression of coronary artery calcification [22,31]. Schlett et al. [24], for instance, found that the epicardial fat volume is nearly twice as high in patients with high-risk coronary lesions as compared to those without coronary artery calcification. Several studies also correlate other cardiovascular risk factors and outcomes to the epicardial adipose tissue volume, such as diastolic filling [16], myocardial infarction [23], atrial fibrillation and ablation outcome [24], carotid stiffness [30], atherosclerosis [17–19], and many others [1,23,28,29,32]. Wei-Ta et al. [33] have also shown that high coronary artery calcium score is associated to a higher general cancer incidence.

Moreover, some studies address the importance of the mediastinal fat and its correlation with pathogenic profiles, health risk factors, diseases such as coronary artery disease and disorders such as metabolic syndrome [27,34,35]. Some associate the mediastinal fat along with the epicardial fat to carotid stiffness [19,30]. Others associate both to atherosclerosis and coronary artery calcification [19,31]. Sicari et al. [2] have also shown how the mediastinal fat rather than the epicardial fat is a cardiometabolic risk marker.

In addition, a 16-year study (that assessed a total of 384,597 patients) associates a rate of approximately 38.4% of death in the subsequent 28 days after a major coronary event [36]. The same study also concludes that fatal cases are slightly less associated to female individuals. Another study ranks cardiovascular accidents as the most common cause of sudden natural death [37]. Therefore, automatically evaluating the amount of fat related to the human heart contributes greatly to avoid such outcomes.

### 2.2. Related works

Some of the first semi-automated segmentation methods for the epicardial fat were proposed around 2005. Coppini et al. [5], for instance, apply a preprocessing step to remove all other structures apart from the heart by using a region growing strategy. Thereafter, an experienced user is required to scroll through the slices to place from 5 to 7 control points along the pericardium border, if visible. Therefrom, Catmull-Rom cubic spline functions are automatically generated to obtain a smooth closed pericardial contour. Finally, the epicardial fat is simply quantified by thresholding, since it is theoretically located within this generated contour. In Pednekar et al. [38], a method for the segmentation of abdominal adipose tissue was proposed. The work of Kakadiaris et al. [39] has further extended the method of [38] to the segmentation of epicardial fats.

Coppini et al. [5] focused on reducing the user intervention. On their method, an expert is still necessary to scroll through the slices between the atrioventricular sulcus and the apex in order to place some control points on the pericardium. The amount of essential points is not clearly described. Nevertheless, the required amount of slices to be analyzed is apparently lesser than the ones required by the method of Dey et al. [40]. Moreover, the authors also present their solution in a 3D space, and claim that Dey et al. [40] do not [5]. The overall focus of their work was to describe their method mathematically. However, the work lacks on evaluation, especially on describing the general accuracy of their method [5].

Barbosa et al. [41] proposed a segmentation for the epicardial fat that was more automated with regard to the ones previously addressed. The authors start using the same preprocessing methodology applied on the work of Dey et al. [40] and further apply a high level step for identification of the pericardium. The identification is done by tracing lines originating from the heart centroid up to the pericardium layer and interpolating the outer points with a spline. Although this approach may be interesting, simple, and highly applicable for virtually any proposed method, the reported results are not impressive. Only 4 out of 40 images were correctly segmented in a fully automatic way.

Shahzad et al. [7] proposed the first fully automated method for epicardial fat segmentation in 2013. Their method uses a multi-atlas approach to segment the pericardium. Their multi-atlas approach is based on registering eight atlases to a target patient and fusing these transformations to obtain the final result. They analyzed 98 patients in their experiments and reported a Dice similarity index of 89.15% to the ground truth and a low rate of approximately 3% of unsuccessful segmentations. Nevertheless, they did not provide information about the overall processing time.

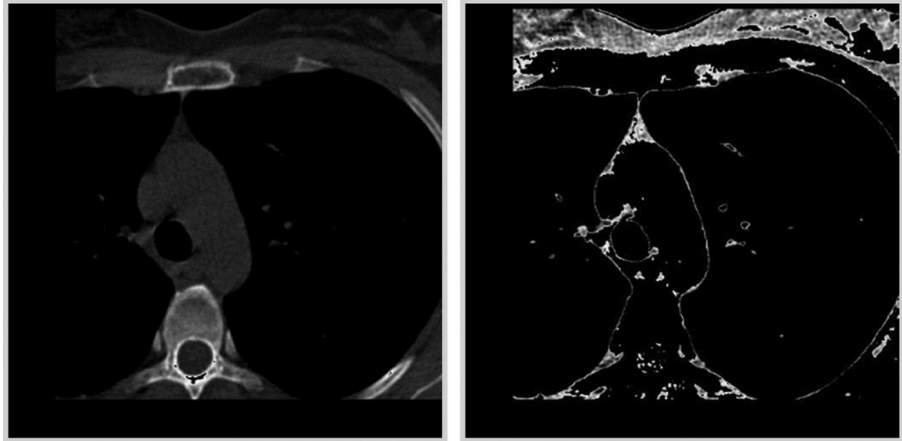

**Fig. 3 – Distinct ranges comparison: from −200 to 1000 HU on the left and from −200 to −30 HU on the right (fat range).**

Ding et al. [8], in 2014, proposed an approach similar to the one of Shahzad et al. [7]. The authors also segment the pericardium using an atlas approach, which consists of a minimization of errors after applying transformations to the atlas along with an active contour method. The reported mean Dice similarity coefficient was equal to 93% and the authors claim that their results were achieved in 60 s in a simple personal computer. Although their segmentation seems to be the most precise in the current literature, the reported computing time is poorly described and too fast for segmenting and transforming an entire cardiac scan, which consists of roughly 50 images. They also present a work [42] that segmented the aorta instead of the pericardium, and compare their achieved time (60 s) to the 15 min of the former [42]. If these 60 s correspond to just the time it takes for the algorithm to minimize the transformations, then this comparison is not feasible. Furthermore, they report that on their approach the atlases' images were pre-aligned to a standard orientation. Therefrom, there is a comparison with only one of these atlases to speed up the process. The remaining pericardium contour follows the pre-aligned pattern, which is a reported limitation. Besides, they did not describe how each one of these atlases is chosen as correct for each possible case.

Although the epicardial fat segmentation has been significantly addressed in the literature, on the other hand, the mediastinal fat has not been addressed as a target for automatic segmentation despite the fact of being a cardiovascular risk marker as previously mentioned in this section.

## 3.    Materials and methods

Computed tomography holds two important advantages over conventional radiographs: three-dimensional image reconstructions and the capability to quantify X-ray attenuation. Attenuation is expressed in CT as Hounsfield Units (HU). The X-ray beams used for diagnostic radiology are composed of photons with a broad spectrum of energies [43]. Molteni [44] explains how HU may be different for a single type of material among distinct CT apparatus or even between the same scanner model if different technical factors are applied such as distinct interpolators. In fact, each system and manufacturer incorporates a unique combination of X-ray source, detector array and projection geometry. Hence, when aiming to segment information stored in HU, these variations should be properly accounted.

In order to properly access the cardiac adipose tissue within CT scans, we need to consider an interval around −100 HU, which, in turn, corresponds to the overall fat of the human body [44]. Coppini et al. [5] and Shmilovich et al. [45] defined the cardiac adipose tissue interval as (−190,−30) while Spearman et al. [9] defined as (−195,−45) and Shahzad et al. [7] as (−200,−30). In this work, we consider the largest proposed interval, which corresponds to the one used by Shahzad et al. This chosen interval fits properly in a 8 bits-depth image and no interpolation is required, avoiding any possible loss of data. Fig. 3 highlights the differences between a CT image (or slice) extracted from a DICOM file on the (−200,1000) HU range and (−200,−30) HU range, respectively. Every pixel that is not within the given ranges is painted black in both occasions.

The overall steps of our methodology are described in Fig. 4. The inputs are CT scans in the DICOM format, where one slice (which should be relatively next to the aorta) of the arbitrary patient is automatically selected for finding the parameters of the registration. After the selection, the slice is converted to a fat-ranged image and is further rescaled to a common empirically chosen scale, standardizing every scan. The parameters of the translation are then computed based on this image already in the fat range. Once the parameters of the translation are found, the entire set of images of the same patient is transformed based on the scaling and translation parameters.

After the registration, twenty randomly chosen patients from Rio de Janeiro were selected to compose a ground truth of cardiac fat. A physician and a computer scientist have manually segmented the cardiac fats of exactly 878 images belonging to those twenty patients. The generated ground truth is available at [46] and its demographic table is shown in Table 1. Thereafter, as highlighted in the third rectangular box of Fig. 4, features extracted from this ground truth were provided as input to several classification algorithms, which generate predictive models. The training process required to generate the predictive model is done a single time only for an arbitrary dataset (our ground truth). That is, when running the proposed methodology, the user already receives a previously built predictive model. Finally, given features extracted from an incoming patient, the predictive model classifies each pixel of each slice as epicardial, mediastinal, both or none. Dilation is then applied to fill some remnant gaps between the pixel classifications. The final segmented images are then used as source to compute the mediastinal and epicardial fat volumes of the patient using a simple interpolation between the segmented pixels of the several slices, while respecting the distance of acquisition between them.

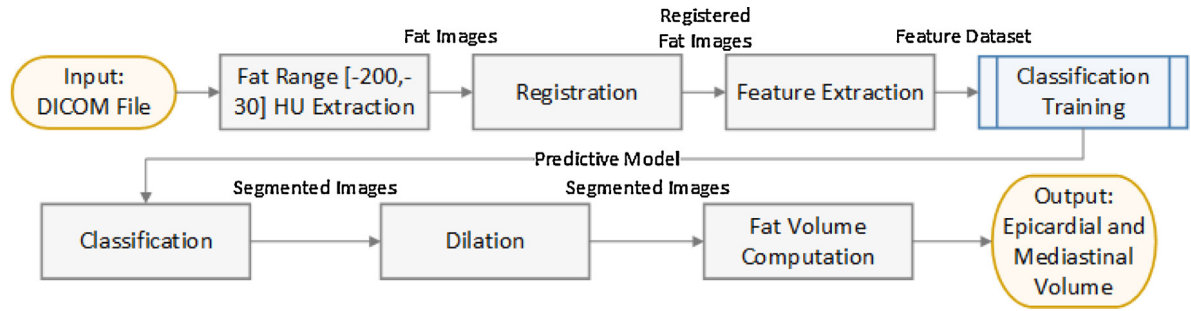

**Fig. 4 – Overall steps of the proposed approach.**

As a remark, this work was approved by the ethics and research committee of Hospital Universitário Clementino Fraga Filho of Universidade Federal do Rio de Janeiro (HUCFF/UFRJ – protocol number 069/10) and of the Brazilian National Institute of Cardiology (protocol number 0324/04-04-2001). All patients were enlightened as to the objectives of this work and were required to sign a consent form.

### 3.1. Registration

The proposed registration comprises two steps: (1) scaling and (2) translation. DICOM files store information related to the real scaling of the patient, regarding the moment of image acquisition [47,48]. In summary, the images of each patient are stored in distinct scales and an attribute of the DICOM file indicates by how much the images should be rescaled in order to transform them back to their natural proportions. Thus, our approach does not consider the scaling issue since it is a trivial operation and, therefore, it focus on the autonomous translation that should be applied to each patient in order to align them. Thus, every slice is rescaled to a common proportion (where the pixel spacing tag should be equal to 0.35 mm) as the very first step.

One may suppose that it is sufficient to apply the rescale operation followed by a translation that centers the rescaled image. However, apart from the variation on the positioning of every patient, when a slice is rescaled and centralized, as Fig. 5 shows, part of the heart may be cropped out in that process. This instance evidences the requirement of an intelligent translation to be applied jointly or after the rescaling.

In translations, it is sufficient to find a single landmark within the image and to use it to displace the image to a standard position. Manual placements of any common landmark, such as several works have proposed [10,38–40], are not eligible here due to the desired autonomous characteristic of our approach. Therefore, the remaining alternative is to automatically find the landmark. In other words, the parameters of our transformation are "searched for" and determined by finding an optimum in the search space of a further addressed objective function.

The proposed translation consists of a combination of (1) a landmark approach and (2) an atlas approach. A common landmark that is relatively easy to recognize is exhibited from the 1st to approximately the 20th slice of every patient, regarding the craniocaudal direction. That common characteristic is denominated retrosternal area and was selected as the landmark to be aligned. The retrosternal area is located on the back of the sternum and is highlighted by the dashed circular spots in the images of Fig. 6 This area does not vary greatly among patients as other cardiac structures do. Besides, it always appears within the boundaries of the CT image, i.e., it is virtually never accidentally cropped off during the acquisition of the cardiac images.

| Table 1 – Patients demographics (ground truth). | | |
|---|---|---|
| Patients | Men: 10 | Women: 10 |
| Ages | Mean: 55.4(±22) | Median: 53 |
| Epicardial fat volume (ml) | Min: 39.2 | Max: 203.4 |
| | Mean: 97.9 | Median: 94.8 |
| Mediastinal fat volume (ml) | Min: 35.6 | Max: 226.2 |
| | Mean: 103.6 | Median: 92.2 |
| Slices | Total amount: 878 | |
| | Per patient median: 42 | Per patient mean: 43.9 |
| Manufacturers | Phillips patients: 9 | Siemens patients: 11 |

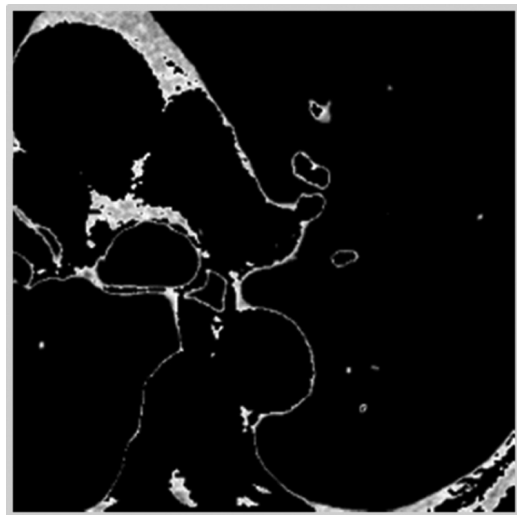

**Fig. 5 – Example of a heart exceeding the boundaries of the image after a rescaling and centralizing translation.**

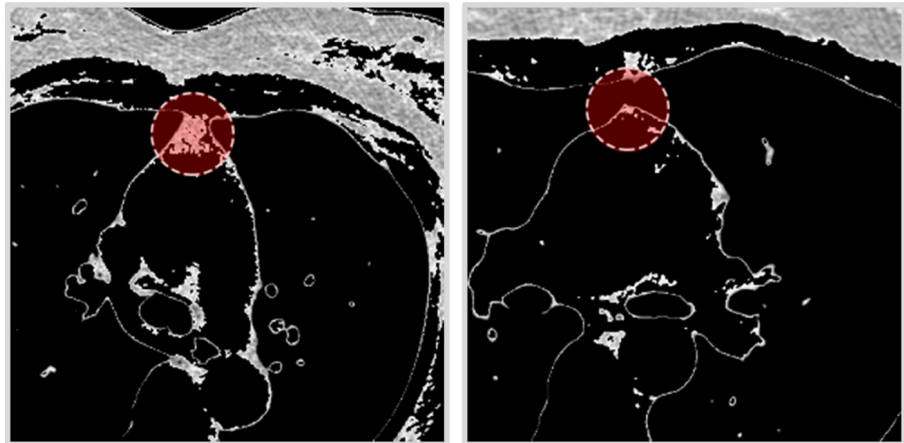

Fig. 6 – Retrosternal area highlighted by the dashed circle.

After establishing the desired landmark, we have randomly selected 10 instances and manually aligned these retrosternal areas in order to compose the atlas, which is used for recognizing the same pattern. Due to the texture variations that occur between distinct manufactures, we chose to threshold each area at gray level 0.2 (assuming the intensities are normalized from 0 to 1) before the arithmetic aggregation of these 10 instances. Fig. 7 illustrates the whole process.

A similarity measure is required in order to assess how much of the atlas is associated to an arbitrary part of a given image. We define as fixed images slices of incoming patients that are to be autonomously segmented. In order to recognize the retrosternal area, the atlas image (also the moving image) is displaced on top of a fixed image and a similarity measure is computed between these two images. We have evaluated the following measures: mean difference (MD), normalized cross-correlation (CC) and mutual information (MI). Furthermore, we have also proposed and evaluated two extensions of MD and MI as similarity measures.

The MD similarity can be obtained from the minimization of Eq. (1), where $M$ denotes the moving image or atlas and $F$ denotes the fixed image (where the retrosternal area should be recognized), $h$ and $w$ stand for the height and width of the atlas, respectively. Furthermore, the $(x, y)$ coordinates stand for every available position for $M$ in $F$ where the upper-left position of the atlas is actually placed. The variables $F_{i,j}$ and $M_{i,j}$ represent the intensity values of the images $F$ and $M$ and $g$ as the exponent of the difference [49].

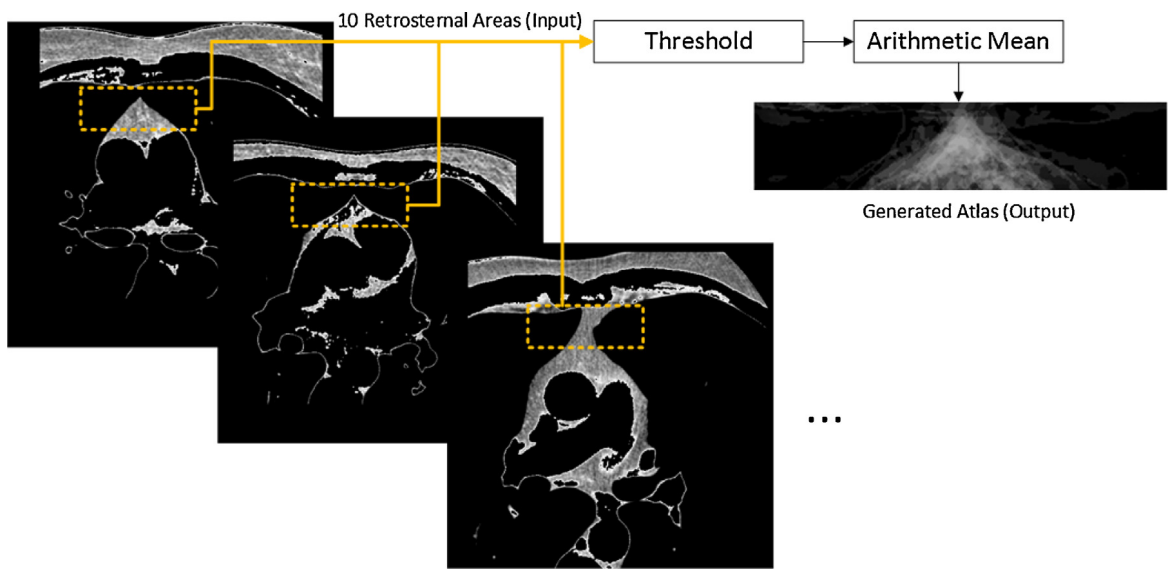

Fig. 7 – Atlas assemblage.

$$MD_{x,y}(F, M, g) = \frac{1}{hw} \left( \sum_{i=y}^{h+y} \sum_{j=x}^{w+x} |F_{i,j} - M_{i,j}|^g \right) \quad (1)$$

The main use for correlation is in the analysis of random-like processes that exhibit similarity in their behavior of occurrence [50]. The equation for the cross-correlation coefficient applied to images is defined in Eq. (2). This equation represents the cross-correlation between the images $F$ and $M$ and their intensity values ($F_{i,j}$ and $M_{i,j}$) The variables $\mu_F$ and $\mu_M$ stand for the mean intensity values of the fixed and moving images, also respectively [49]. The maximal value that $CC$ can achieve is 1, which would imply that the images are in alignment [49]. This measure allows the registration of objects whose intensity values are related by a linear transformation [51].

$$CC_{x,y}(F, M)$$
$$= \left| \frac{\sum_{h+y}^{i=y} \sum_{w+x}^{j=x} (F_{i,j} - \mu_F)(M_{i,j} - \mu_M)}{\sqrt[2]{\sum_{h+y}^{i=y} \sum_{w+x}^{j=x} (F_{i,j} - \mu_F)^2 \sum_{h+y}^{i=y} \sum_{w+x}^{j=x} (M_{i,j} - \mu_M)^2}} \right| \quad (2)$$

Mutual information is a measure of how well one image explains the other [52]. When two images are composed by entirely distinct pixel values, mutual information is zero. The maximum value of mutual information is 1. The mutual information coefficient is defined as $MI$ in Eq. (3), where $m$ is equal to the pixel values within the atlas image $M$ and $f$ is equal to every pixel value of the fixed image $F$ that is within the atlas image boundary, i.e., between the top-left corner of the atlas and the right-bottom corner. The variables $\rho_F$ and $\rho_M$ are the marginal probability distributions, corresponding to intensities in the images $F$ and $M$ respectively [49]. Furthermore, $H(K)$ stands for the Shannon entropy of the histogram $K$ [52] and $\rho_{FM}$ stands for the joint probability.

$$MI_{x,y}(F, M, g) = H(F) + H(M)$$
$$- H(F, M) \left( \sum_{f \in F} \sum_{m \in M} \rho_{FM}(f, m) \log_g \frac{\rho_{FM}(f, m)}{\rho_F(f)\rho_M(m)} \right) \quad (3)$$

In the traditional formulation of the mutual information, each event or object specified by $(f, m)$ is weighted by the corresponding joint probability $\rho_{FM}(f, m)$ This assumes that all objects or events are equivalent apart from their probability of occurrence. However, in some applications, it may be the case that certain objects or events are more significant than others, or that certain types of associations are more important than others [53]. Assuming that the similarity of $E = \{9, 5, 2\}$ and $R = \{2, 9, 5\}$ is being measured by the $MI$ equation, the comparisons between $E$ and $E$ or $E$ and $R$ would yield the same result. However, in our problem, it is clear that the images $E$ and $E$ are more similar than $E$ and $R$ although they hold the same mutual information. Thus, we propose a combination of the mean difference measure and mutual information, originating a weighted mutual information (WMI) measure shown in Eq. (4).

$$WMI_{x,y}(F, M, g)$$
$$= \left( \sum_{f \in M \to F} \sum_{m \in M} \frac{1}{|f - m| + 1} \rho_{FM}(f, m) \log_g \frac{\rho_{FM}(f, m)}{\rho_F(f)\rho_M(m)} \right) \quad (4)$$

The factor $1/(|f - m| + 1)$ defines the applied weight to the equation. When $f = m$, each parcel of the sum of the $MI$ equation is multiplied by 1. In this case, the maximum value of WMI is still 1 and that coefficient is achieved when the images are the same and are perfectly aligned.

Moreover, we also propose an extension of the MD measure called hybrid mean difference (HMD), with the premise of solving issues introduced by MD (such as the non-weighted computing of pixels) by computing the weight of darker and brighter pixels differently while reducing errors that possibly occur. At first, by thresholding the atlas image on the level $t$, we separate two main parts of the image: (1) the one with darker pixels and (2) the one with brighter pixels. Thereafter, the brighter area of the atlas has its pixel values $m$ inverted and subtracted of the values $f$ from the fixed image. If the difference between $f$ and $m$ is higher than 0 then this value is summed to compose a partial error score $S_p$. Furthermore, the pixel values of the moving image that belong to the darker area are subtracted of the fixed image to compose a partial image $I$. The sum of the pixel values in $I$ composes the remaining partial error score $S_n$. The final hybrid difference is given by $S_n - S_p$. By minimizing HMD, we maximize $S_p$ and minimize $S_n$ altogether. When both images $F$ and $M$ are identical, HMD returns 0. However, when regarding distinct images, it returns negative values if $S_n < S_p$.

In summary, by applying the HMD measure we are reducing the error caused by faint pixels of the moving image. When we average images to build an atlas, some of its parts are almost completely dark (value $\cong 0$), and these pixels count almost as black pixels in the traditional MD measure, causing a significant error. In the HMD measure we reduce the weight of these kind of errors. The HMD measure is given by Eq. (5), where $t$ stands for the threshold level, $F$ stands for the fixed image, $M$ stands for the moving image, $(x, y)$ or the position of $M$ within $F$. Furthermore, $g$ stands for the exponent of the difference between the pixel values of $F$ and $M$. Besides, we consider that the intensity values $F_{i,j}$ and $M_{i,j}$ are normalized.

$$HDM_{x,y}(F, M, g, t)$$
$$= \sum_{i=y}^{h+y} \sum_{j=x}^{w+x} \begin{cases} -\max \left( \left( F_{i,j} - \frac{1}{M_{i,j}} \right)^g \cdot 0 \right), & \text{if } M_{i,j} > t \\ |F_{i,j} - M_{i,j}|^g, & \text{otherwise} \end{cases} \quad (5)$$

In addition, to be applied along with the chosen similarity measure, we propose a heuristical confirmation method to reinforce the recognition. The proposed heuristic is straightforward and is summarized as follows. Given a small area $A$ of pixels at the central position of the atlas (the center of the recognized retrosternal area), there should be two points $p_l$ and $p_r$ that continuously move through fat pixels on the left-bottom and right-bottom orientations until they reach convergence. A fat pixel is defined as a pixel that is not background (i.e., black).

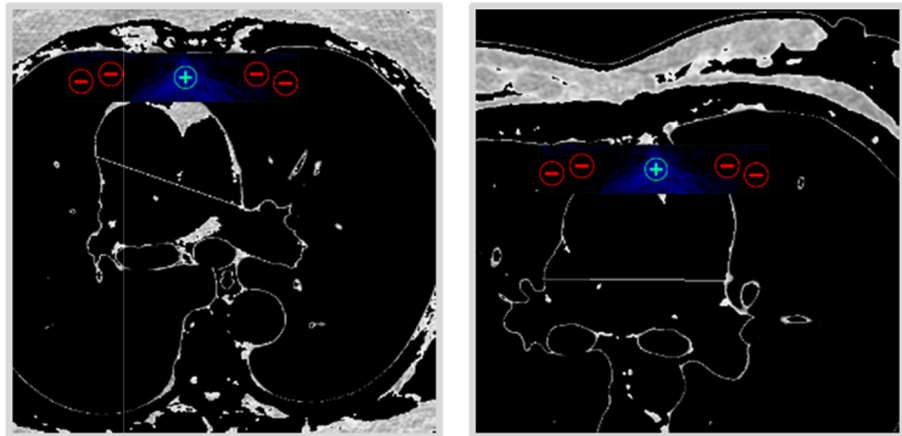

**Fig. 8 – Binding of the two points $p_l$ and $p_r$.**

The thin slanted white lines in the instances of Fig. 8 illustrate the binding of, or the distance between these two points $p_l$ and $p_r$ after their convergence. The action of moving these points on the image is just part of the confirmation method. The logical conditions that reinforce if the selected position for the atlas is correct are: (1) the traced line must be within a certain width ($x$-axis) and (2) both points must also be within a certain distance from each other and from the starting point as well. If this confirmation fails, the settlement of the atlas and the confirmation method must be redone jointly.

The reason for considering an area $A$ instead of a single point is because if a single point is selected, not always are the two points $p_l$ and $p_r$ able to move through fat pixels. This is due to the fact that not every patient has a straight continuity of fat deposits in the retrosternal area, as evidenced by the arrows in the instances of Fig. 9.

Thereafter, from every pair of pixels within $A$ there must be at least one pair $p_l$ and $p_r$ that satisfies the conditions in Algorithm 1 in order to confirm the retrosternal area location. The variables $x_{min}$ and $x_{max}$ stipulate the minimum and maximum distances that the points $p_l$ and $p_r$ must move. The parameter $c$ stands for the convergence threshold, i.e., how many times both coordinates $p_l$ and $p_r$ should move. The parameter $u$ stands for the maximum amount of background pixels that can be skipped at each movement of the coordinates $p_l$ and $p_r$. The function $d(p_1, p_2)$ returns the discrete Euclidian distance in pixels between the points $p_l$ and $p_r$. Finally, it is important to highlight again that $p_l$ and $p_r$ are required to move only on top of fat pixels (i.e., those that are not background).

Algorithm 1. Confirmation method

| | |
|---|---|
| 1 | confirmationMethod($x_{min}, x_{max}, c, u$) |
| 1.1. | *For each combination of point $p_l$ and $p_r$ within $A$;* |
| 1.1.1. | *Place $p_s$ at the center of $p_l$ and $p_r$;* |
| 1.1.2. | *For $c$ times:* |
| 1.1.2.1. | *Try to move $p_l$ and $p_r$ by $1\ldots u$ pixels;* |
| 1.1.3. | *If $\|p_l \cdot x - p_r \cdot x\| > x_{min}$ and $\left\| p_l \cdot x - p_r \cdot x \right\| < x_{max}$* |
| 1.1.3.1. | *If $d(p_l, p_s) > d(p_r, p_s)/2$ and $d(p_r, p_s) > d(p_l, p_s)/2$* |
| 1.1.3.1.1. | *If $d(p_l, p_s) > x_{min}$ and $d(p_r, p_s) > x_{min}$* |
| 1.1.3.1.1.1. | *Return true;* |
| 1.2. | *Return false;* |

In summary, what the heuristic in Algorithm 1 essentially does is, starting at the initial coordinates of $p_l$ and $p_r$, it attempts to displace these points only if there is an available fat pixel to move next, respecting a set hierarchy. The $p_r$ stands for "right point or pixel", which infers that $p_r$ must move down and rightwards, whereas $p_l$ must move down and leftwards. The coordinates of $p_l$ and $p_r$ are changed only when the next moveable pixel is a fat pixel. Moreover, the hierarchy of displacement must respect the following statements: (1) at first, each point must be moved on the horizontal direction (left or right), (2) if the associated pixel is not a fat pixel then it returns to the original position and must be moved on the vertical direction (down only). Finally, (3) if the pixel associated to the previous movement is also not a fat pixel then the point returns to the original position and is moved diagonally (left or right along with a downward movement).

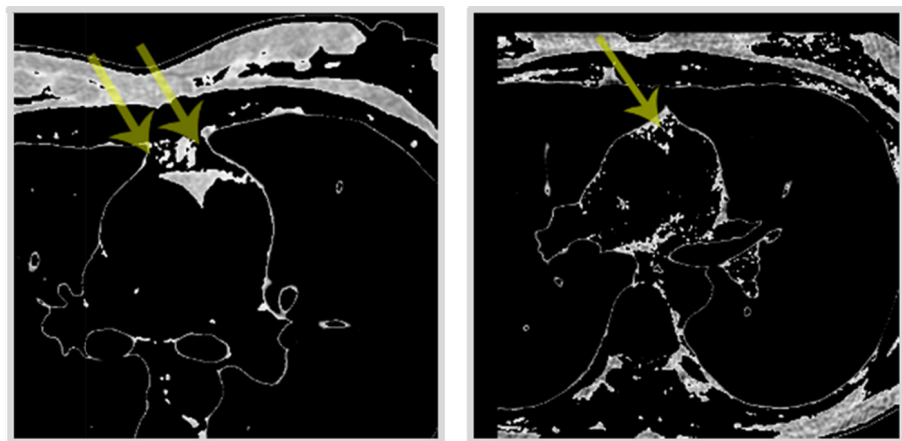

**Fig. 9 – Lack of continuous fat deposits in the retrosternal area.**

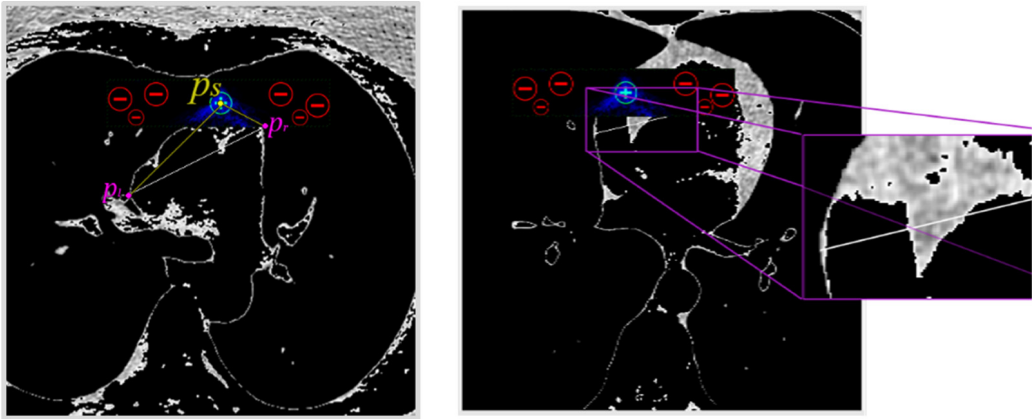

**Fig. 10 – Logical conditions of the confirmation method.**

That process of choosing which direction to move and verifying if the associated pixel is fat is done at most $u$ times for each iteration of $c$. The $u$ variable was introduced to overcome some interpolation problems introduced by the resizing operation. Although we have applied just the bicubic interpolation, sometimes, the resultant image contained small gaps between pair of pixels that were previously continuous. Therefore, the use of the $u$ variable was essential. Finally, that whole process of moving the coordinates of $p_l$ and $p_r$ done at most $c$ times until the algorithm converges.

Finally, the Boolean condition in the line 1.1.3 of Algorithm 1 verifies if the heart is within certain width. The condition in line 1.1.3.1, on the other hand, verifies if both coordinates of $p_l$ and $p_r$ moved, to some extent, equally to each other. This condition was coined to avoid situations such as the one depicted by the first instance in Fig. 10, where the distance from $P_s$ (starting point) to $P_r$ is significantly lesser than the distance from $P_s$ to $P_l$. In a similar fashion, the condition in line 1.1.3.1.1 verifies if both points are at least at a meaningful distance from the initial point $P_s$ in order to avoid situations depicted by the second instance in Fig. 10, where $p_l$ and $p_r$ are very close to the atlas image. It is clear that the absence of these conditions produces some unsuccessful recognitions.

In summary, the retrosternal area is firstly recognized using the atlas and a similarity measure. After this recognition, the confirmation method is run in that recognized position. If the confirmation method returns false then the recognition step is run all over again. However, in this rerun of the recognition step, at each evaluated position for the atlas, the confirmation method is also trigged jointly. Therefore, the autonomously recognized landmark is defined as the position that minimizes the similarity measure and that also returns true on the confirmation method. These described steps define our objective function. Based on the recognized landmark, the images are then translated to a standard position.

### 3.2. Classified segmentation

Considering that the slices to be segmented are already registered, the proposed methodology for segmentation is applied, which involves classification algorithms. The classification approach consists of three main steps: (1) extracting features from the ground truth, (2) training the predictive model on the extracted data and (3) further classifying an incoming CT scan with the generated predictive model where the classes of the pixels are unknown. The steps (1) and (2) do not need to be redone every time a slice needs to be classified. In fact, if that were true, the method would take so long to converge that it would be unpractical. The generated dataset is very big and, therefore, the steps (1) and (2) are very slow processes. Thus, the step (3) is independent of (1) and (2) once they have already been concluded. The user application starts directly at step (3). These three mentioned steps are represented by the three containers named Step 1 to 3 in Fig. 11, which is a more detailed extension of Fig. 4 regarding the classification process.

The step 1 consists of iterating every pixel of the images that compose the ground truth and extracting features related to the iterated pixels. The feature vector of each pixel represents a line on the extracted dataset and, therefore, various pixels of the ground truth images compose the final dataset that is used to train predictive models for further segmentation. There is a huge variety of features that can be extracted from images; some of them are specifically valuable for certain types of images and arrangement of data. Later, we define and select some features that are extracted for our problem. Moreover, we decided not to normalize the features after the extraction. The main reason is so that without normalization, we are able to generate various predictive models from different training sets and aggregate these predictive models with reduced effort.

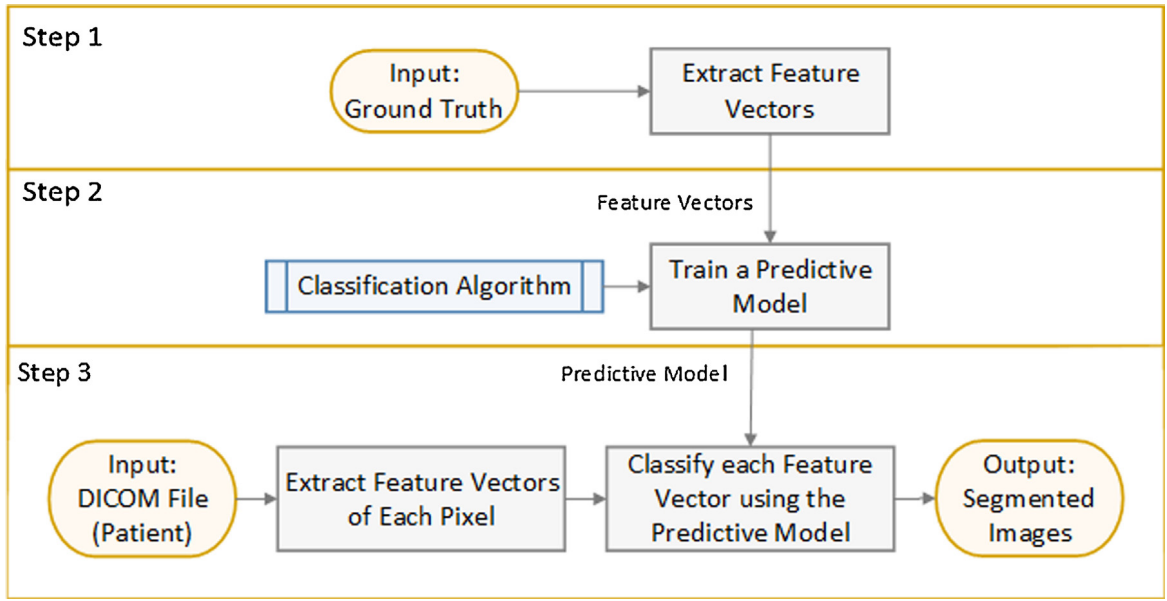

**Fig. 11 – Overall steps required for the classified segmentation.**

The steps 2 and 3, i.e., the steps to be applied after extracting the features, are totally related to the classification algorithms. On this aspect, the incoming pixel can be assigned to at most 3 different classes (mediastinal fat, epicardial fat and pericardium, or none of them, which is not viewed as an actual class). The images of the ground truth had their epicardial, mediastinal fats and a transitional area between the two (that can be qualified as the pericardium) segmented in different colors (red, green and blue, respectively) such as the instance in Fig. 12. Furthermore, it should be reminded that black pixels (0) are set as background and are therefore not classified by the predictive model.

Some classification algorithms work only with binary classes. Thus, we divided the 3 possible classes into a binary mapping for each class. When the pixel does not belong to any of the three classes it is considered background. The classes are disposed in three columns in the dataset; each column represents a class and receives true or false as value. When training the algorithm for the epicardial fat, the columns that represent the remaining classes (mediastinal and pericardium) should be removed prior to the training in order to avoid predictive models trained regarding these classes as normal attributes.

The features we have selected for extraction are divided in three main categories: (1) the primary features, which are directly related to the pixel information, (2) the secondary, which are related to the image or to a neighborhood window and (3) the tertiary, which are related to data that was already derived from the image or from a neighborhood window. Thus, the possible primary features that were extracted are the pixel value along with its $x$ $y$ and $z$ coordinates. For secondary features we have extracted the $x$ and $y$ coordinates of the pixel with respect to the center of gravity of the image ($x - x_g$, $y - y_g$) where ($x_g$, $y_g$) is obtained with Eq. (6). Besides, a $\bar{i}x\bar{j}$ neighborhood of pixels around the iterated pixel $P_{y,x}$ was considered for extracting information, where $\bar{i} = \bar{j} = 2q + 1|q > 0$. If any pixel of this neighborhood extrapolates the boundaries of the image then it is handled as a black pixel. From this neighborhood, the following secondary features were extracted: (1) a simple arithmetic mean of the gray values, (2) the geometric moments $M(0,1)$, $M(1,0)$ and $M(1,1)$, obtained with Eq. (7) and, (3) a proposed coefficient of smooth variation (CSV) of the gray values.

$$(x_g, y_g) = \left( \frac{M(1,0)}{M(0,0)}, \frac{M(0,1)}{M(0,0)} \right) \tag{6}$$

$$M(m, n) = \sum_i \sum_j j^m i^n P_{i,j} \tag{7}$$

When the entire neighborhood window is regarded, the convolution of the CSV is given by Eq. (8), where $\beta$ is a constant and should be adjusted to avoid overflows. In our approach, we have considered $\beta$ to be $10^7$. The weight $\sqrt[4\infty+1]{\beta}$ is multiplied by the pixel value $P_{i,j}$ of the window. For explanatory purposes, considering a $7 \times 7$ neighborhood window, the groups of distances of the Gaussian Kernel is 9, as opposed to 3 of the proposed CSV. This means that CSV is faster to compute and is relevant in a real-time scenario.

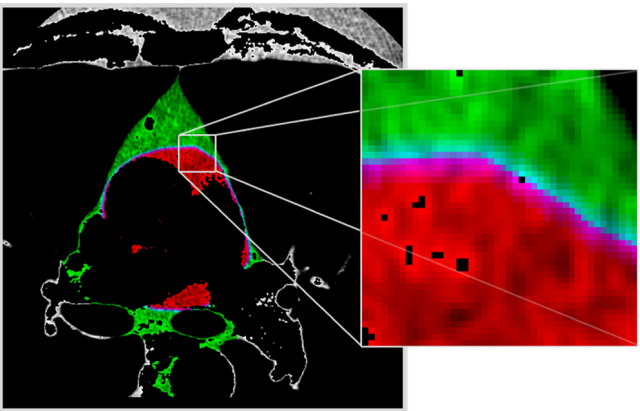

**Fig. 12 – Ground truth definition.**

| Table 2 – Extracted features and used parameters. | | |
|---|---|---|
| Type | Feature | Parameters |
| Primary | Gray value | – |
| – | $x$, $y$ | – |
| – | $z$ (slice number) | – |
| Secondary | $x$ relative to the center of gravity | – |
| – | $y$ relative to the center of gravity | – |
| – | Arithmetic mean | – |
| – | Coefficient of smooth variation | – |
| | Geometric moments | $M(0,1)$, $M(1,0)$, $M(1,1)$ |
| Tertiary | Moments of the co-occurrence matrix | $(\Delta x, \Delta y) = \{(0, 1), \ (1, 0), \ (1, 1)\}$ combined to $\{M_1, M_2, M_3, M_4\}$ |
| | Run percentage | $\theta = \{0°, 45°, 90°, 135°\}$ |
| | Grey level non-uniformity | $\theta = \{0°, 45°, 90°, 135°\}$ |

$$CSV = \sum_i^{\bar{i}} \sum_j^{\bar{j}} {}^{d_\infty+1}\sqrt{\beta} P_{i,j} \qquad (8)$$

Furthermore, moments $M_{1...4}$ at the distances (0,1), (1,0) and (1,1) of the co-occurrence matrix were extracted as tertiary features. The reason for selecting those distances is so that there are images where the outer layer of the heart is of approximately one or two pixels of thickness. Moreover, based on the run length matrix, the run percentage and gray level non-uniformity were also extracted as tertiary features respecting the set of orientations $\theta = \{0°, 45°, 90°, 135°\}$. All these tertiary features were obtained from the neighborhood window. The extracted features are summarized in Table 2 along with the applied parameters, a total of 31 features were initially extracted from the images.

Texture is considered an important regional descriptor for segmentation and classification of various types of medical images. Thus, in this case, the reason for extracting texture-based features (most of the secondary and tertiary features) was mainly due to the hypothesis that the epicardial and mediastinal fat yield a slightly difference on their texture that can be partially accounted by the analytical process of these features.

The co-occurrence matrix associates the number of co-occurrences of a gray level of a pixel $p_a$ to a gray level $p_b$ on an image $P$ at a given distance $(\Delta x, \Delta y)$ for all the $p_{1...n} \in P$ where $n$ is the total amount of distinct gray values in $P$. The function $C_{\Delta x, \Delta y}(p_a, p_b)$ in Eq. (9) denotes the co-occurrences between the pixel values $p_a$ and $p_b$ at a distance $(\Delta x, \Delta y)$ [54].

$$C_{\Delta x, \Delta y}(p_a, p_b) = \sum_i \sum_j \begin{cases} 1, & if \ P_{i,j} = p_a \ and \ P_{i+\Delta y, j+\Delta x} = p_b \\ 0, & otherwise \end{cases} \qquad (9)$$

The probability of a gray level $p_a$ co-occurring with $p_b$ at a certain distance is given by the number of times these two gray values co-occur divided by all the co-occurrences of every pair of gray values $(k, l)$ at the same distance. Hence, the probability of $p_a$ co-occurring with $p_b$ is given by the function $P_{\Delta x, \Delta y}(p_a, p_b)$ in Eq. (10). The moment of the co-occurrence matrix is given by the function $M_g$ in Eq. (11), where $g$ is the moment degree [54].

$$P_{\Delta x, \Delta y}(p_a, p_b) = \frac{C_{\Delta x, \Delta y}(p_a, p_b)}{\sum_k \sum_l C_{\Delta x, \Delta y}(k, l)} \qquad (10)$$

$$M_g = \sum_k \sum_l P_{\Delta x, \Delta y}(k, l)(k - l)^g \qquad (11)$$

The run length matrix follows a similar principle in relation to the run length encoding. Given a direction $\theta$, the number of identical gray values of an image in that same direction can be accounted to compose a matrix. Thus, one axis of this matrix represents all the possible gray values comprised in the image and the other axis represents the length of the run, i.e., all the possible times the gray values appear continuously on the direction $\theta$. The four possible directions for $\theta$ are $0°$, $45°$, $90°$ and $135°$, where $0°$, for instance, represents the horizontal direction. If we are to compute the matrix for the $0°$ direction, all the lines of the image are accessed and when any gray level $p$ is continuously repeated $l$ times, the value of the run length matrix at the position $(p, l)$ is increased by one. We assume that the run length matrix for an arbitrary image is $R_\theta$, where $R_\theta(p, l)$ represents the times that the gray value $p$ appears continuously $l$ times in the image respecting the direction $\theta$. Thus, the gray level non-uniformity feature is computed by solving Eq. (12). The summations are done for every possible $p$ and $l$ of the matrix [54].

$$G_\theta = \frac{\sum_p \left( \sum_l R_\theta(p, l) \right)^2}{\sum_p \sum_l R_\theta(p, l)} \qquad (12)$$

The run percentage consists of summing all the elements of the run length matrix and dividing it by the area $S$ of the image, as shown in Eq. (13). The areas of the neighborhood window are constants relative to the same dataset and, therefore, that division is irrelevant in this case. Nevertheless, to maintain the usual formulation we kept the division [54]. The sum of the run length matrix appears to be a relevant feature

| Type | Feature | Epicardial | Mediastinal | Pericardium |
|------|---------|-----------|-------------|-------------|
| **Table 3 – Ranked evaluation of features per class.** | | | | |
| Primary | Gray value | 1 | – | – |
| – | $x$ | 2 | 1 | – |
| – | $y$ | 3 | 2 | – |
| – | $z$ (slice number) | – | – | 19 |
| Secondary | $x$ relative to the center of gravity | 4 | 3 | 1 |
| – | $y$ relative to the center of gravity | 5 | 4 | 2 |
| – | Arithmetic mean | 6 | – | 4 |
| – | Coefficient of smooth variation | – | 5 | 3 |
| – | Geometric moments | 14, 15 | 11 | 15, 16, 17, 18 |
| Tertiary | Moments of the co-occurrence matrix | 7, 8, 9, 10 | 6, 7 | 5, 6, 7, 8, 9, 10, 11 |
| – | Run percentage | 13 | 8, 9, 10 | – |
| – | Gray level non-uniformity | 11, 12 | – | 12, 13, 14 |

since it could show how uniform is an arbitrary part of the image.

$$R_\theta = \frac{\sum_p \sum_l R_\theta(p, l)}{S} \qquad (13)$$

After extracting all these features from the ground truth we have used the linear forward selection as search method along with the attribute subset evaluator to rank the proposed features on their importance for each one of the classes of our problem. Table 3 compares the position of features in the ranking on the epicardial, mediastinal and pericardium columns (the lower the better). The features that are not described in this table were not selected as sufficiently relevant by the algorithm. The ranking of the features varies substantially depending on which class we are analysing. That fact was already expected and is one of the reasons we have considered this long range of features for extraction, in order to propose a satisfactory unified method.

For the epicardial fat, the best parameters for the moments of the co-occurrence matrix were distance (0,1) and $g = 4$, followed by distance (1,0) and $g = 3$. For the mediastinal fat the best set of parameters were distance (0,1) and $g = 4$, followed by distance (1,1) and $g = 4$. Moreover, for the epircardium, the distance (0,1) and $g = 1$ followed by (0,1) and $g = 2$ were the most valuable. With regard to the features on the run length matrix, the orientations $0°$ and $90°$ were the most valuable for the epicardial fat, where $90°$ was the only valuable orientation for the run percentage feature. As to the mediastinal, the most valuable orientations were $45°$, $90°$ and $135°$, in this order. However, for the pericardium, the most valuable orientations were, respectively, $0°$, $45°$ and $135°$. Furthermore, the run percentage feature was not significant for the pericardium class due to the fact that it appears just as a small contour of fat around the heart where the run length matrix cannot extract much valuable information. As it can be seen, it is extremely difficult to set the parameters in a non empiric fashion due to the low contrast of the fat textures.

Moreover, a few features were counter-intuitively ranked. For instance, $z$ is basically irrelevant among all of the three classes. However, theoretically, it appears to be a valuable feature since the arrangement of the types of fat varies greatly from one slice to the other (on the $z$ index). Nevertheless, if another algorithm for the attribute selection was regarded in this analysis, maybe the results would change significantly with relation to this feature. However, extracting $z$ is a cheap computational operation, so that it would not impact the overall processing time of the algorithm. Another surprise was the ranking of the run percentage and gray level non-uniformity features, both are based on the run length matrix but none could be more significant than the moments of the co-occurrence matrixes in any of the three occasions.

The coefficient of smooth variation performed much better than the arithmetic mean on the case of the mediastinal fat, the mean arithmetic feature was not even considered relevant. The CSV performed better than the arithmetic mean on the pericardium as well. However, it performed worse than the pixel value and the arithmetic mean on the epicardial fat. Nevertheless, we tend to think that repeating this evaluation with another algorithm would at least slightly change that panorama. However, it is important to highlight that repeating this experiment with other algorithm was not done due to the time required to processes those attribute selection algorithms.

The tertiary features are unarguably the most expensive to be extracted. Thus, based on the feature ranking and computational cost, we have removed the features based on the RLM, and removed the geometric moments as well. The geometric moments were considered relevant in all of the three cases but in any of them they were significantly important for the predictive model creation. Moreover, some set of parameters that performed worse than others were removed as well. In conclusion, from the set of 31 initially proposed features and parameters we have discarded 16 of them, resulting in 15 extracted features. These 15 features, along with their parameters, are shown in Table 4. The impact of this reduction is discussed in Section 6.

## 4.    Experiments and partial results

Successful registrations were defined as trials where the algorithm-chosen retrosternal landmark coincided with the actual retrosternal area of the fixed image, regarding slightly positioning variations within this area (max of 30 pixels). These autonomous recognitions were evaluated as successful by two observers. A total amount of 82 randomly chosen slices were provided as fixed images to the atlas-driven recognition regarding the five similarity measures previously addressed in Section 3. The results obtained by each measure are shown

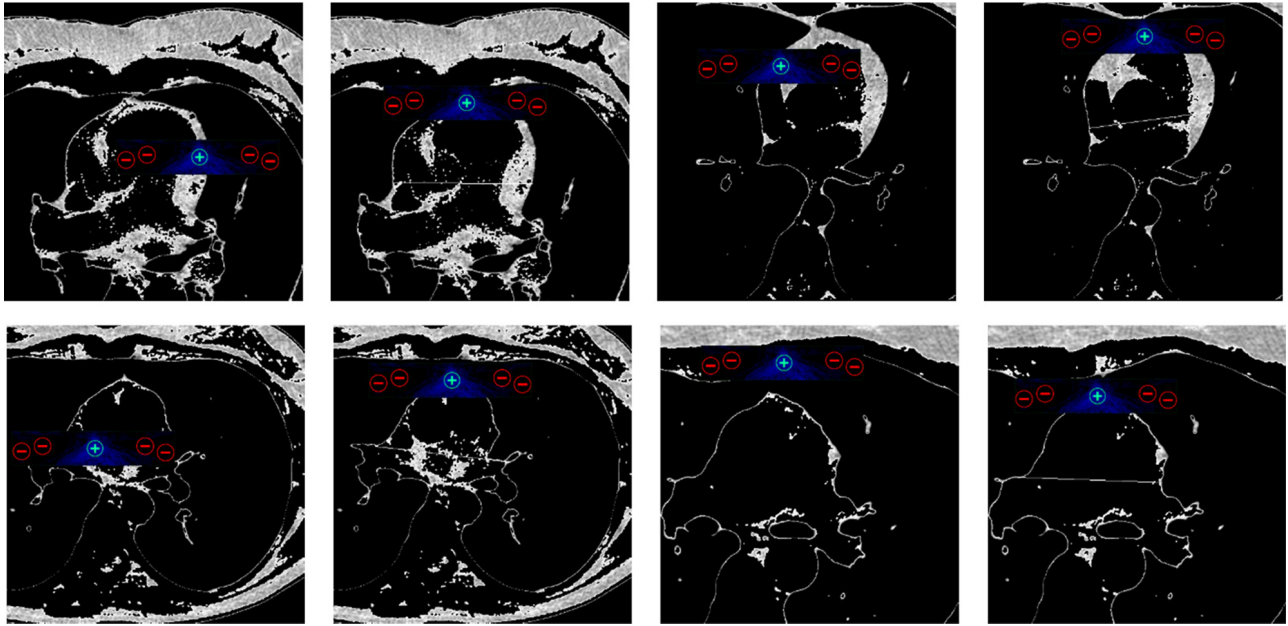

Fig. 13 – Recognitions without (left) and with (right) the confirmation method.

**Table 4 – Empirically selected features and parameters (more valuable).**

| Type | Feature | Parameters |
|---|---|---|
| Primary | Gray value | – |
| – | x, y | – |
| – | z (slice number) | – |
| Secondary | x relative to the center of gravity | – |
| – | y relative to the center of gravity | – |
| – | Arithmetic mean | – |
| – | Coefficient of smooth variation | – |
| Tertiary | Moments of the co-occurrence matrix | (0,1) combined to $M_{1...4}$ (1,1) combined to $M_{2...4}$ |

in Table 5, where the values are the average between the two observers.

In short, the proposed HMD performed better than the normalized cross-correlation, mutual information and better than the proposed weighted mutual information as well. The highest rate of successful recognitions achieved with HMD was equal to 95% against 73% of the weighted mutual information measure. In other words, once the HMD was applied, the proposed recognition method was able to successfully locate the retrosternal area in 78 instances out of a total of 82.

**Table 5 – A comparison of the employed similarity measures (* proposed measures).**

| Measures | Successful recognitions (%) |
|---|---|
| Hybrid mean difference (HMD)* | 95 |
| Weighted mutual information (WMI)* | 73 |
| Mean absolute difference (MD) | 66 |
| Mutual information (MI) | 64 |
| Normalized cross-correlation (CC) | 58 |

However, to improve this recognition even more and to avoid future unpredicted occasions such as deformations in the retrosternal area, we have proposed and applied a confirmation method, as previously addressed. Consequently, the confirmation method improved the rate of successful recognitions from 95% to 100% among the 82 tested slices. In approximately 95% of the time (rate of successful recognitions of the HMD), the confirmation method would be run only once and that does not significantly increase the overall processing time. On the other 5%, the confirmation method is run at every evaluation of positions in F. Therefore, in this case, the increase on the processing time is fairly noticeable. However, when applying some heuristics to select just the most suitable positions for evaluation, the additional increase in time shrinks to just about 2 min in a simple personal computer. As a matter of reproducibility, the set of empirically chosen parameters for the confirmation method was: $c = 0.6W$, $u = 0.003(W + H)$, $x_{min} = 0.2W$, $x_{max} = 0.55W$ and the size of the area $A = (013W, 0.04H)$ where $W$ and $H$ are the width and height of the fixed image, with respect to Algorithm 1 and assuming that the CT images are 512 × 512 pixels wide. Fig. 13 illustrates the four occasions (among 82) where the absence of the confirmation method leads to unsuccessful recognitions and their respective corrections on the right side. That is, the images should be analyzed pairwise, where the right instance is the corrected version.

Accuracy is defined as the sum of the true positive and true negative occurrences divided by the total population. As previously addressed, we are dealing with three distinct binary classes in this work. As from here, when only one accuracy rate is provided, we define it as being the arithmetic mean of the three classes of our work (epicardial, mediastinal and pericardium). For the classification experiments in this work we have employed the Weka library [55]. Weka is an open-source collection of machine learning algorithms maintained by the University of Waikato.

**Table 6 – Accuracies and convergence time for a single patient.**

| Algorithm | Accuracy % (66% split) | Accuracy % (10× CV) | Training time (s) | Classification time (s) |
|---|---|---|---|---|
| RandomForest | 98.9 | 99.0 | 97.57 | 63.43 |
| J48 | 98.8 | 99.0 | 143.17 | 101.83 |
| J48Graft | 98.8 | 99.0 | 203.83 | 130.17 |
| REPTree | 97.6 | 98.7 | 13.28 | 13.68 |
| RandomTree | 97.6 | 98.0 | 12.26 | 15.14 |
| SPegasos | 92.4 | 92.4 | 143.18 | 115.89 |
| HyperPipes | 92.3 | 91.4 | 0.15 | 6.10 |
| RBFNetwork | 85.8 | 85.7 | 247.57 | 193.43 |
| DecisionStump | 85.6 | 85.5 | 18.74 | 17.43 |
| NaiveBayes | 81.9 | 81.8 | 7.74 | 16.39 |

In order to evaluate convergence speed and to reduce the amount of classification algorithms to be experimented we have extracted the proposed features of a single patient of the ground truth, producing approximately 250,000 feature vectors at total. The time that each algorithm took to construct the predictive model based on two thirds of the patient data as well as the time it took to evaluate the model on the remaining one third (66% split) were specifically taken into consideration. It is important to reduce the computational time spent on training to shrink the time consumption of the whole analysis. Furthermore, it is not possible to predict exactly if an algorithm will converge soon or later given a passed time limit. That is, if we set the time limit to 250 s, there may be some algorithms that converge in 260 s but others that converge in hours or days. Certainly this time limiting is not fair, however, there are not many alternatives as to this extent, and many algorithms have to be evaluated. We have set the time limit to 250 s because some of the best rated algorithms in the literature such as the RandomForest, J48 and NaiveBayes converged within that limit and we consequently defined it as sufficient. Besides, an experiment that evaluates the arguably optimum size for the neighborhood window is further performed, and these algorithms have to be rapid in their training phase. Moreover, the parameters experiments also consume a significant time, where the training phase of each algorithm has to be run several times.

Therefore, for that single patient of the ground truth, we have tested all the classification algorithms present in Weka on its version 3-6-11. Some of these algorithms are: the Support Vector Machine (SVM), Sequential Minimal Optimization (SMO), Naïve Bayes, Radial Basis Function Network (RBFNetwork), Random Trees, C4.5 (or J48), Primal Estimated Sub-Gradient Solver for SVM (SPegasos), REPTree, k-Nearest Neighbors (kNN or IBk), Multilayer Perceptron and others. Among all the tested algorithms, we have selected for further analysis the ones that converged within 250 s. The parameters of each algorithm were based on their standards with some adjustments and the best result was selected. On this evaluation, the size of the neighborhood was 25 × 25. The comparison of the results is shown in Table 6, the time values and the first accuracy column are referent to a 66% random-selected split, while the second accuracy column is referent to a 10-fold cross-validation (CV). Both test modes were mainly considered to show that there is not much variation between them since the total amount of instances in the dataset is huge.

At first, the REPTree algorithm appears to be the best choice since it achieved a great accuracy in a relatively fast convergence time (training and classification) if compared to the remaining. In addition, the result of HyperPipes was rather interesting. In this case, it returned a very good accuracy and converged very rapidly. However, when applied to a bigger dataset and a set of images of distinct patients, the results obtained through the HyperPipes algorithm start to differ drastically.

In order to diminish unfair comparisons of classifiers we have tested the classifiers in Table 6 along with a variation on the neighborhood size. Hypothetically, one may consider that some classifier performs better on a neighborhood of certain size. Fig. 14 shows some charts that represent the accuracy of each classifier on the y-axis and the variation of the neighborhood size in pixels on the x-axis. These neighborhoods are squared neighborhoods, that is, they are based on the sup metric instead of the Euclidian one. The accuracies are divided per class in three charts. Since there is a huge amount of possible neighborhood sizes, these accuracies were achieved using a 66% random split method as test mode from the data of 20 patients. However, some slices were skipped from the processing to decrease the convergence time (those who have an odd index). The extracted features composed datasets of approximately 1.5 gigabytes for each neighborhood size that were provided to the classifiers. The period to train and evaluate the model lasted, in some cases, up to 25 min for each combination of neighborhood size and classification algorithm.

The five classifiers in Fig. 14 were the ones who performed the quickest on this large dataset (20 patients). The HyperPipes was the fastest and always converged virtually within 0.15 s but, in this case, the best accuracy that it could achieve was around 70%. We can state by this evidence that this algorithm is not generalizable. In other words, it fails to generate a general predictive model from the moment that more than one patient is regarded for the training. However, it is a rather simple algorithm and, in fact, the achieved low accuracy was somehow supposed to happen.

The REPTree classifier did not converge remarkably faster than the remaining decision tree algorithms on this large dataset, even though fast convergence is probably the main characteristic of the REPTree. RBFNetwork was faster than RandomForest and SPegasos was slower but both returned lower accuracies. RandomTree and DecisionStump were just a little faster than RandomForest and REPTree but returned significantly lower accuracies and, therefore, were disregarded due to the massive presence of decision tree algorithms on the previous convergence time experiment. J48Graft returned similar accuracies if compared to RandomForest but its convergence was approximately 1.4 times slower and, due to that matter, it was impracticable to evaluate all the neighborhood sizes for this classifier. Naive Bayes and J48 took more time to converge on this large dataset than SPegasos and, therefore, they could not be precisely evaluated.

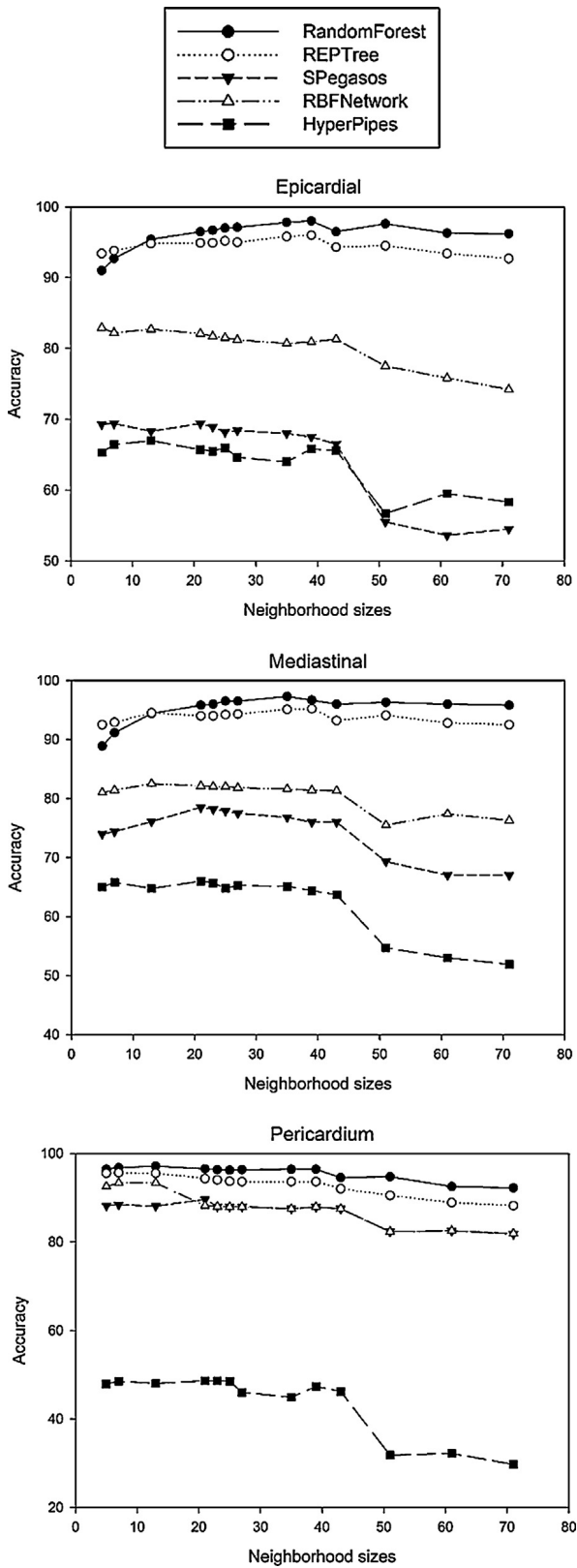

**Fig. 14 – Accuracies of classifiers versus neighborhood sizes.**

In conclusion, the accuracy of virtually every classifier started to slightly decrease after the size of $25 \times 25$ pixels and to sharp decrease after $39 \times 39$. In fact, the larger the neighborhood size becomes, the more it converges to the size of the image, producing more loss of information. This behavior was expected, however, that analysis was performed in order to set an arguably perfect size for the neighborhood window, for posteriorly refining the results. Moreover, the size of the neighborhood window is proportional to the time the algorithm takes to extract features from the image and, consequently, to converge. Therefore, by evaluating this tradeoff we decided that $25 \times 25$ was the most suited size for the neighborhood window.

Finally, to complement our work we have analyzed the overfitting degree along with the produced visual aspects of the RandomForest, REPTree, J48, J48Graft, SMO and SPegasos algorithms. Overfitting is defined as a lack of generalization or a super adjustment to the training dataset induced by classification algorithms during the training phase. For instance, a algorithm that achieves high accuracies evaluating its predictive model on the training set but fails to get high accuracies on similar datasets can be considered overfitted to the training data [56]. Our overfitting analysis consisted of purposely training classifiers with data of a single patient and using the generated predictive model to segment another patient scan of a different manufacturer. The classifier who visually performed better, i.e., the one that segmented the images more similarly to the ground truth was defined as having the lowest overfitting. The parameters for the features were the same for all the classifiers and the parameters of the classifiers were empirically chosen. Results of a single instance segmented by the addressed classification algorithms are shown in Fig. 15. For this trial, we have disregarded the blue layer (i.e., the pericardium or the unknown area).

We concluded that the RandomForest was the algorithm that overfitted the least. In fact, its segmentation result is the most similar if compared to the ground truth in Fig. 15. REPTree, J48 and J48Graft algorithms performed a less sparse segmentation but it is clear that their predictive model was heavily induced by the features derived from the spatial disposition of the pixels ($x$ and $y$ coordinates) and not much on the texture. That fact is evidenced by the lines and columns on the segmentation produced by these classifiers.

The SMO and SPegasos, both algorithms based on the SVM, outputted the segmented image entirely in yellow. The yellow color is a combination of the colors red and green. Therefore, what the algorithm essentially did was classifying all the bright pixels as epicardial and mediastinal fats. Thus, we can infer that the SMO and SPegasos algorithms lack heavily on generalization, at least for this evaluation. They took an easy path assuming that every pixel is both epicardial and mediastinal fat. This may be associated to the weakness of the classifier in a sense that, if the generated model does not predict well, then it is better to just assume that everything is true or everything is false to raise the accuracy by some margin.

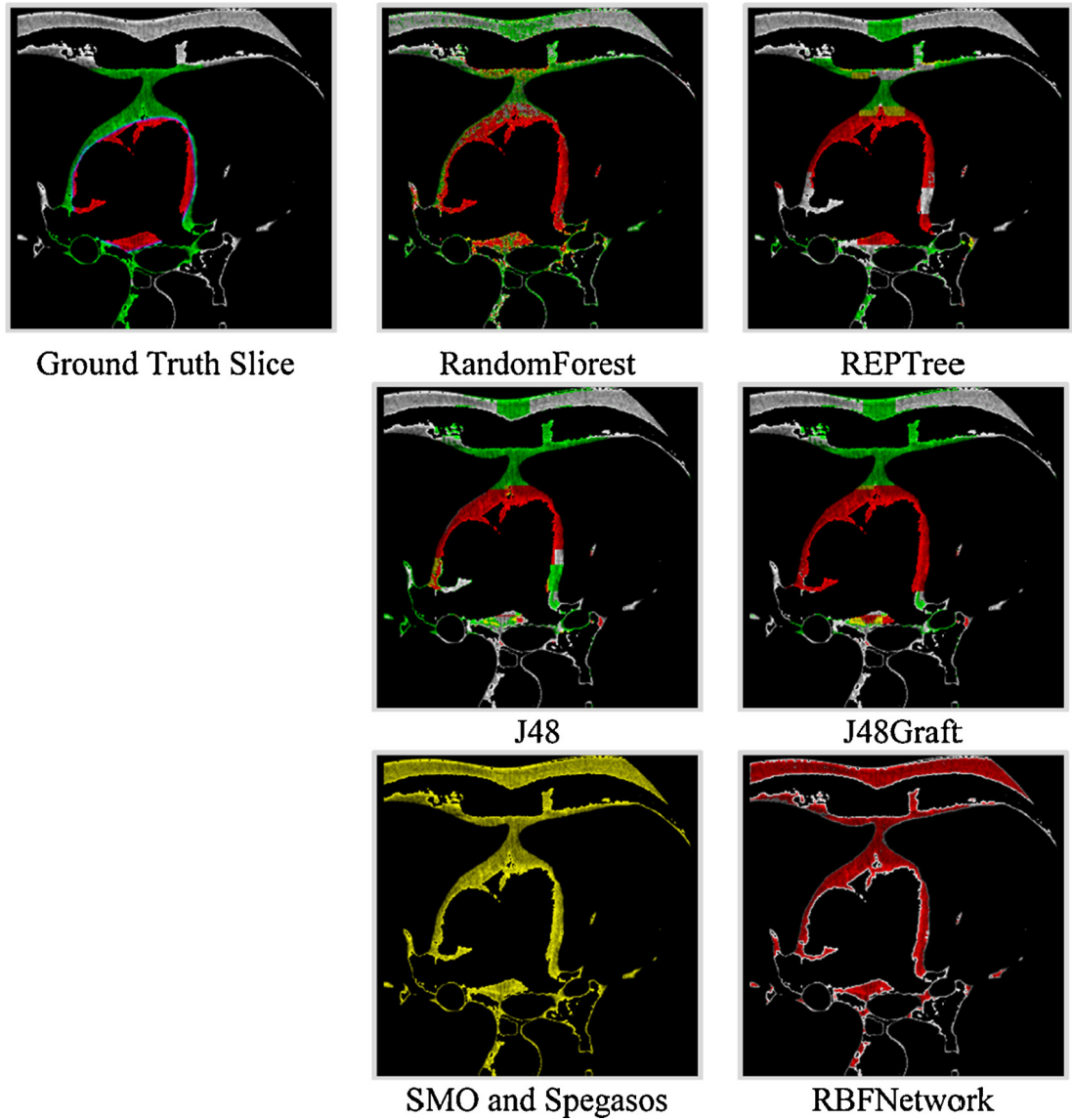

**Fig. 15 – A ground truth slice segmented by some algorithms trained on another patient.**

The predictive model generated by RBFNetwork was similar to SMO and SPegasos to some degree. With regard to the mediastinal fat (green), the algorithm classified every pixel as false. On the other hand, regarding the epicardial fat (red), it classified almost everything but the outline as true. The result of the RBFNetwork segmentation is a little worse than assuming that every pixel belongs to both classes, but is still being a bad result.

Among the decision tree algorithms, we can say that the RandomForest was the one that least overfitted, directly followed by the J48Graft algorithm. The J48Graft converges in a slower fashion if compared to the RandomForest and produces similar accuracies but a very different type of visual segmentation (less sparse). J48Graft converged faster than the J48 algorithm and produced better segmentations and accuracies. Thus, J48 and REPTree were the ones that most suffered overfitting between these four decision tree algorithms.

## 5. Results

The slices in each line of Fig. 16 belong to 4 distinct patients and represent some results of the proposed registration. The images on the left column are in the (−200,1000) HU range and the ones on the right column are in the fat-range (−200,−30) HU. The images of patients 1 and 2 were acquired with the same scanner. Although the texture between these two seems not to change, there is a divergence on the scale and a slight divergence on the positioning. The third and fourth patient data were acquired with other two distinct scanners. Hence, in total, these 8 images were obtained from three distinct manufactures. The images on the right column are much more aligned between themselves and within a corresponding scale than the ones shown on the left column. The registration reduces the error of the segmentation, speeds up the volume quantification and may even fasten and simplify the generated predictive model.

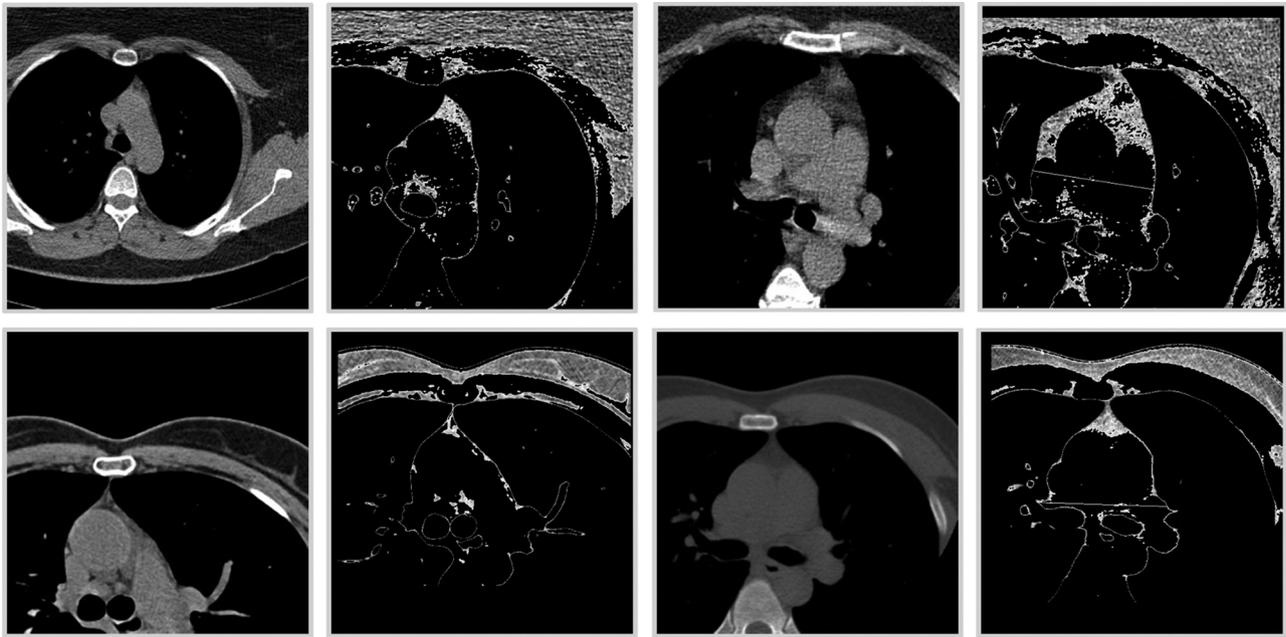

Fig. 16 – Four intersubject registered patients.

**Table 7 – Obtained results using the Random Forest algorithm on 10-fold cross validation.**

| Class | Accuracy % | TP rate % | TN rate % | FP rate % | FN rate % | Avg. ROC area | Kappa |
|---|---|---|---|---|---|---|---|
| Epicardial ■ | 98.5 | 98.9 | 98.3 | 1.1 | 1.7 | 0.985 | 0.966 |
| Mediastinal ■ | 98.4 | 97.2 | 98.8 | 2.8 | 1.2 | 0.984 | 0.956 |
| Pericardium ■ | 96.4 | 85.2 | 98.0 | 14.8 | 2.0 | 0.964 | 0.835 |

As to the extent of the classification algorithms, Random-Forest was the classifier that performed the best if the speed, accuracy, overfitting and visual segmentation analyses are considered. To refine our results we produced a dataset of approximately 2.5 gigabytes originated from the patients of the ground truth. However, in this occasion, no slice was skipped during the feature extraction. This extracted dataset is also available directly on the Weka's arff format at [46]. For this occasion, we considered the 10-fold cross validation as test mode. Table 7 contains the accuracies and the confusion matrixes of RandomForest with standard parameters using the chosen neighborhood of 25 × 25 pixels.

This work is the first to propose the use of classification algorithms for cardiac fat segmentation. The nature of the classification algorithms implies that they are tightly related to confusion matrixes and to the accuracy index. In fact, this is usually the standard way to evaluate the performance of a classification. However, just one of the three main related works evaluates their segmentation on the basis of accuracies and confusion matrixes. Table 8 compares the results of these three main related works. The values of cells that were left blank were not provided by the authors. The segmentation proposed by Kakadiaris et al. [39] is semi-automated, while the works of Shahzad et al. [7] and Ding et al. [8] are fully automated. All these three works proposed methods for segmenting just the epicardial fat and, therefore, we compare just our epicardial fat segmentation in Table 8.

**Table 8 – Comparison of the epicardial fat segmentation.**

| Work | TP rate % | ROC area | Dice index % |
|---|---|---|---|
| Kakadiaris et al. [39] | 85.6 | – | – |
| Shahzad et al. [7] | – | – | 89.15 |
| Ding et al. [8] | – | – | 93.0 |
| This work | 98.9 | 0.985 | 98.1 |

Fig. 17 visually compares two automatically segmented slices (right column) of two distinct patients to their ground truth (left column). The predictive model used to segment both was the same, trained on the data of other 20 patients (ground truth). It is important to highlight that the blue color is not present on the automatically segmented images due to the fact that it was interpreted as a transitional area between the epicardial and mediastinal fats. Thus, the pericardium classification was slightly different than the other two. That is, if a pixel was classified as red (epicardial) or green (mediastinal) before being classified as blue (pericardium), then it remains at its previously classified color. Otherwise, the pixel is painted yellow (both epicardial and mediastinal) instead of blue.

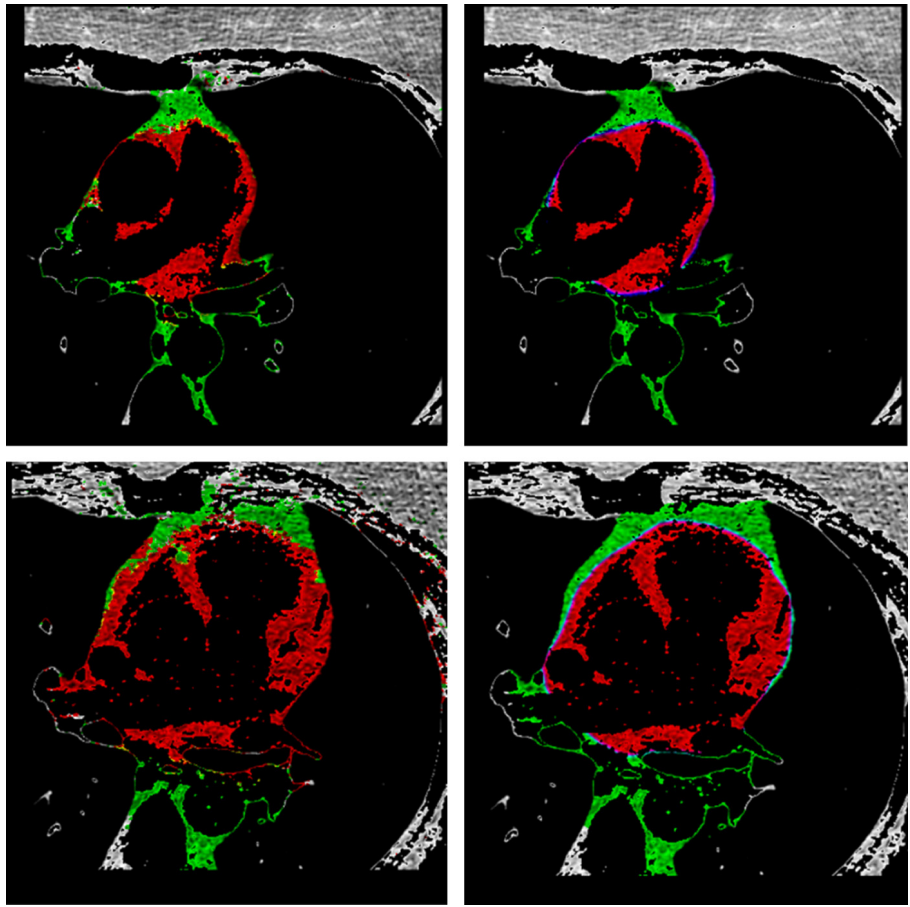

**Fig. 17 – Comparison of two automatically segmented slices (left) and their ground truth (right).**

## 6.     Conclusion

Our methodology achieved the best results for the autonomous epicardial fat segmentation in the current literature. Furthermore, we are the first to propose a unified segmentation method for both types of cardiac fats and the first to autonomously segment the mediastinal fat. We define the term "unified method" as using a unique methodology to segment the epicardial as well as the mediastinal fats. Moreover, we are also the first to provide a public available cardiac fat ground truth for further comparisons. In addition, we corroborate on the appliance of classification algorithms to image segmentation.

As long as the right features are selected, the appliance of classification algorithms to image segmentation is highly prone to success and may surpass many usual segmentation methods on several aspects. RandomForest is often rated as the best decision-tree algorithm and proved its efficiency in our analysis. We also concluded that decision tree algorithms provided much better performance over neural networks, function-based classification algorithms and probabilistics models such as Naive Bayes (where the maximum achieved accuracy was 77.1%). The mean accuracy achieved with RandomForst for the epicardial and mediastinal fats was 98.5% (99.5% if the features are normalized) with a mean true positive rate of 98.0%. The mean Dice similarity index was 97.7%. Every registration was considered successful by two observers and every segmentation could also be considered successful in a sense that no major error or unpredicted behavior occurred.

The dataset used in our work regarded CT images produced by different scanners from mainly two different manufacturers (Siemens and Philips). We did not include features related to the scanner model and manufacturer into the prediction to avoid several repetitions in the extracted dataset, which would significantly increase its size. However, we believe that if these features were regarded or if distinct predictive models were generated for each scanner model and manufacturer the results would improve even more. Moreover, ensemble methods use multiple learning algorithms to obtain better performance [57]. In this work, we have used a "single" classification algorithm to segment the images instead of a combination of them. For instance, the J48Graft could be combined to RandomForest (though RF is already an ensemble) and perhaps to REPTree algorithms to increase even more the overall accuracy of the predictive model.

Besides, we did not consider normalizing the extracted data due to three main factors. The non-use of normalization implies: (1) distinct predictive models can be concatenated easily, (2) the extracted features from a unclassified pixel do not need to be converted to the normalized range, which would imply extra considerations on the concatenation of distinct predictive models, and (3) the raw values of each feature permit a simpler and directly readable evaluation of the data. Nevertheless, we normalized the extracted data before applying the RandomForest algorithm just to check by how much its indexes would improve, as they usually do when normalized. Thus, as previously addressed, the RandomForest had its accuracy improved, in average, up to 99.5% with the normalization.

In conclusion, the achieved results are very satisfactory. Still, the current approach requires an adaptation in order to be applied in real time. With the feature selection, we have reduced the amount of features to 15 (instead of 32 initially proposed) and in this case we have been able to process a patient scan in 1.8 h in contrast to almost 1 day in the former occasion. This feature selection improved the true positive indexes of the mediastinal and epicardial fats but reduced the pericardium accuracy (the blue layer). The experiments in this work were mostly run in an Intel octa-core CPU with 8 Gb of RAM and no dedicated graphics card. As future works we want to migrate this methodology or part of it to the GPU in order to be able to process this information in real-time. Furthermore, we are also confident that the proposed methodology can be adapted to other imaging modalities with little effort.

## Acknowledgements

É. O. Rodrigues wants to thanks CAPES for the financial support. A. Conci wants to thank the CNPq project 302298/2012-6.